\documentclass[Physsubmission, Phys]{SciPost}
\usepackage{amsmath,amsthm,mathtools}
\usepackage{physics}
\usepackage{axodraw4j}
\usepackage{tikz}
\DeclareMathOperator{\ep}{\epsilon}
\DeclareMathOperator{\FF}{{}_3F_2}
\DeclareMathOperator{\F}{{}_2F_1}

\DeclareMathOperator{\Euler}{e^{2\gamma_E\epsilon}}

\newcommand{\G}[1]{\Gamma\left({#1}\right)}

\newcommand{\f}[4]{\F\left(#1,#2;#3;#4\right)}

\newcommand{\ff}[6]{{}_3F_2\left(#1,#2,#3;#4,#5;#6\right)}

\hypersetup{
    colorlinks,
    linkcolor={red!50!black},
    citecolor={blue!50!black},
    urlcolor={blue!80!black}
}

\usepackage[charter]{mathdesign}
\DeclareSymbolFont{usualmathcal}{OMS}{cmsy}{m}{n}
\DeclareSymbolFontAlphabet{\mathcal}{usualmathcal}
\numberwithin{equation}{section}
\begin{document}

\begin{center}{\Large \textbf{
The Diagrammatic Coaction and Cuts of the Double Box\\
}}\end{center}

% TODO: write the author list here. Use initials + surname format.
% Separate subsequent authors by a comma, omit comma at the end of the list.
% Mark the corresponding author with a superscript *.
\begin{center}
Aris Ioannou\textsuperscript{$\star$}and
Einan Gardi\textsuperscript{$\star\star$}
\end{center}

\begin{center}
{\bf } Higgs Centre for Theoretical Physics, School of Physics and Astronomy\\
The University of
Edinburgh, Edinburgh EH9 3FD, Scotland, UK
\\
* a.ioannou-5@sms.ed.ac.uk\\
** Einan.Gardi@ed.ac.uk
\end{center}

\begin{center}
\today
\end{center}

% For convenience during refereeing (optional),
% you can turn on line numbers by uncommenting the next line:
%\linenumbers

\definecolor{palegray}{gray}{0.95}
\begin{center}
\colorbox{palegray}{
  \begin{tabular}{rr}
  \begin{minipage}{0.1\textwidth}
    \includegraphics[width=35mm]{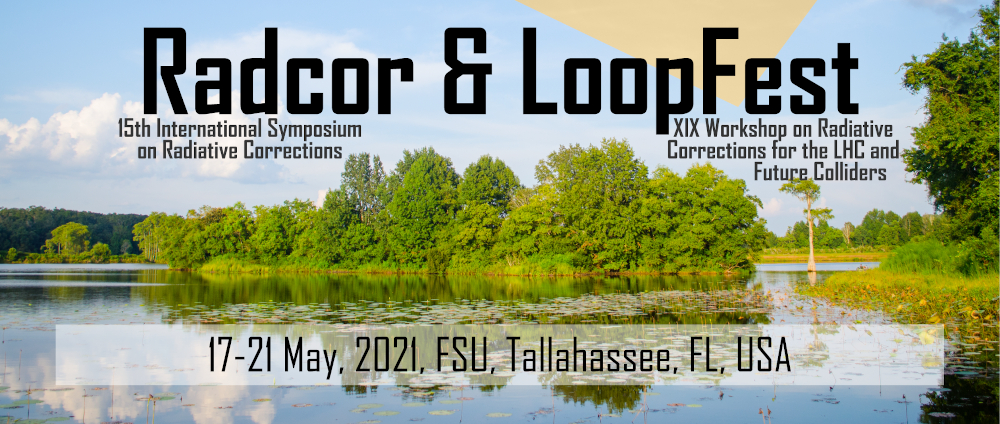}
  \end{minipage}
  &
  \begin{minipage}{0.85\textwidth}
    \begin{center}
    {\it 15th International Symposium on Radiative Corrections: \\Applications of Quantum Field Theory to Phenomenology,}\\
    {\it FSU, Tallahasse, FL, USA, 17-21 May 2021} \\
    \doi{10.21468/SciPostPhysProc.?}\\
    \end{center}
  \end{minipage}\end{tabular}}\end{center}

\section*{Abstract}
{\bf
% TODO: write your abstract here.
The diagrammatic coaction encodes the analytic structure of Feynman integrals by mapping any given Feynman diagram into a tensor product of diagrams defined by contractions and cuts of the original diagram. Feynman integrals evaluate to generalized hypergeometric functions in dimensional regularization. Establishing the coaction on this type of functions has helped formulating and checking the diagrammatic coaction of certain two-loop integrals. In this talk we study its application on the fully massless double box diagram. We make use of differential equation techniques, which, together with the properties of homology and cohomology theory of the resulting hypergeometric functions, allow us to formulate the coaction on a range of cuts of the double box in closed form.}

\vspace{10pt}
\noindent\rule{\textwidth}{1pt}
\tableofcontents\thispagestyle{fancy}
\noindent\rule{\textwidth}{1pt}
\vspace{10pt}
\section{Introduction}
\label{sec:intro}
It has been proposed \cite{2017} that dimensionally regulated Feynman integrals can be endowed with a diagrammatic coaction involving contractions and cuts of subsets of propagators. This idea was first established in the context of one-loop Feynman integrals \cite{Abreu:2017mtm}. As one-loop Feynman integrals are always expressible in terms of multiple polylogarithms (MPLs)\cite{Duhr:2012fh}, the diagrammatic coaction could be formulated and checked order by order in the dimensional regulator $\ep$, using the established coaction on the MPLs \cite{2001math......3059G}.

The hope is, however, that this algebraic structure extends to all Feynman integrals. A further step in this direction was facilitated by the formulation of a coaction on generalized hypergeometric functions, which are expandable order by order in $\ep$.

It has been shown \cite{brown2020lauricella},\cite{Abreu:2019wzk}, that the "local coaction" which applies to the MPL appearing in the Laurent expansion of hypergeometric functions, such as those obtained upon evaluating Feynman integrals in dimensional regularization, is compatible with the "global coaction" on these integrals. The global coaction is defined using the integral representation of hypergeometric functions for fixed $\ep$, without reference to their expansion. This formulation has been recently used \cite{Abreu:2019wzk} to generalize the diagrammatic coaction beyond one loop, and explicitly construct the coaction for a variety of two-loop Feynman-integral topologies \cite{Abreu:2021vhb}.

 In the present talk we report on further progress in applying the diagrammatic coaction at two loops, by considering the on-shell double box topology. The choice of topology is motivated by the fact that the double box has a higher number of propagators than those of the previously considered examples, which, when contracted, result in a higher variety of subtopologies which appear in the diagrammatic coaction.
 
 This double box integral is not yet known in a closed  form in $\ep$, but we nevertheless are able to compute its cuts in a closed form, through differential equation techniques, and construct the diagrammatic coaction of these cuts, thus providing further insight into the form of the two-loop diagrammatic coaction.
\section{Background}
\subsection{The double box Integral}
The massless double box Feynman integral, as well as the integral representation of its subtopologies, can be expressed in the following form:
	\begin{equation}\label{Int}	I\left(D,s,t,a,b,\{n_i\}\right)\scalebox{0.96}{$=$}\frac{e^{2\gamma_E\ep}}{\left(i\pi^{\frac{D}{2}}\right)^2}		\int_{-\infty}^{+\infty}
\scalebox{0.96}{$\frac{d^Dk\; d^Dl\;\left((k-p_3)^2\right)^a\left((l-p_1)^2\right)^b}{\left(k^2\right)^{n_1}\left(l^2\right)^{n_2}\left((l+k)^2\right)^{n_3}\left((k+p_1)^2\right)^{n_4}\left((l+p_3)^2\right)^{n_5}\left((k+(p_1+p_2))^2\right)^{n_6}\left((l-(p_1+p_2))^2\right)^{n_7}}$}
\end{equation}
where $D$ is the spacetime dimension, $\ep$ the dimensional regularization parameter, and $a$, $b$ and ${n_i}$ are non-negative integer-valued exponents.
We may view the double box integral, as well as any Feynman integral, as a pairing $I\left(D,s,t,a,b,\{n_i\}\right)=\bra{\gamma}\ket{\omega_{a,b,\{n_i\}}}$ of an element of a cohomology group $\ket{\omega_{a,b,\{n_i\}}}$, the differential form being integrated, with an element of a homology group $\bra{\gamma}$, the contour of integration.

\paragraph{The cohomology group.}
In the case of the massless double box the cohomology group can be generated by the differential form:
\begin{equation}\label{difform}\footnotesize
		\ket{\omega_{a,b,\{n_i\}}}=\frac{e^{2\gamma_E\ep}}{\left(i\pi^{\frac{D}{2}}\right)^2}\frac{d^Dk \wedge d^Dl\;\left((k-p_3)^2\right)^a\left((l-p_1)^2\right)^b}{\left(k^2\right)^{n_1}\left(l^2\right)^{n_2}\left((l+k)^2\right)^{n_3}\left((k+p_1)^2\right)^{n_4}\left((l+p_3)^2\right)^{n_5}\left((k+(p_1+p_2))^2\right)^{n_6}\left((l-(p_1+p_2))^2\right)^{n_7}}
\end{equation}
 The case where all $n_i\neq0$ corresponds to the double box topology itself, while setting any of the $n_i$ parameters to zero is equivalent to considering a subdiagram with the corresponding propagator contracted.  By virtue of integration-by-parts (IBP) relations, one may generate a minimal set differential forms which spans the space of integrands associated with any particular topology. In the case of the double box this basis is eight-dimensional, consisting of six independent master integrands, corresponding to subtopologies, and two master integrands corresponding to the double box top topology, for which we choose $n_i=1$ for all $i$.
 
 The inequivalent master integrands of the top topology are generated by the numerator insertions $\left((k-p_3)^2\right)^a$ and $\left((l-p_1)^2\right)^b$. The necessity of numerator insertions  is a feature that appears at two loops and beyond, where, in contrast to the one-loop case, it is no longer sufficient  to consider subsets of the top topology propagators with unit powers for obtaining a master integrand  basis. This consideration complicates the generalization of the coaction to the two-loop level.

\paragraph{The homology group.}
The homology group is generated by all the inequivalent integration contours that prescribe the integration path for the loop momenta. The usual Feynman-integral contour prescribes a path which takes every component of the loop momenta over the whole range $\left(-\infty,+\infty\right)$. One may modify the integration contour by allowing it to encircle any of the poles of the propagators. By virtue of the residue theorem, this translates into placing the encircled propagators on-shell, which is also referred to  as "cutting" the propagators. Cut Feynman diagrams are represented  with the on-shell propagators featuring a cut line.

At one loop, the homology group for a particular class of diagrams is generated by all the non-empty subsets of the propagators put on shell. Starting from two loops, a contour cannot be characterized only by the set of propagators it encircles; there are more independent choices of contours encircling the same set of propagators, just as there are more than one master integrals associated with the same topology. In the case of the double box this translates to two inequivalent maximal cuts, the same number as master integrands of the top topology.
\paragraph{The duality condition.}
For many Feynman integrals it is possible to make a basis choice for the homology and the cohomology group generators such that the $\ep$ expansion of the period matrix $\bra{\gamma_i}\ket{\omega_j}$ is:
\begin{equation}\label{dual}
\bra{\gamma_i}\ket{\omega_j}=\delta_{ij}+\mathcal{O}(\ep).
\end{equation}
Equation \eqref{dual} is called the duality condition. Choosing the generators such that they satisfy the duality condition is essential for the coaction to take the form outlined in the following section. The duality condition will also serve as a boundary condition for the cuts computed by differential equations in section \ref{difsex}.

\subsection{The coaction on integrals}
The coaction on integrals \cite{2017} takes the general form:
\begin{equation}\label{diagcoa}
\Delta\left[\bra{\gamma}\ket{\omega}\right]=\sum_{i} \bra{\gamma}\ket{\omega_i} \otimes \bra{\gamma_i}\ket{\omega}.
\end{equation}
 The left hand side of the equation features the coaction applied on an arbitrary integral resulting from any differential form $\ket{\omega}$ integrated along a contour $\bra{\gamma}$. In the right hand side, the original contour stays constant on the left entry of the tensor product, while the original differential form stays constant on the right entry. The sum is performed over dual pairs of basis elements, where the integrand $\ket{\omega_i}$ determines the left entry while its corresponding dual contour $\bra{\gamma_i}$ the right entry.
\subsection{Hypergeometric functions}
A coaction based on the integral representation of the hypergeometric functions has been constructed in \cite{abreu2019diagrammatic}. The simplest hypergeometric function is the Gauss $\F$ which admits the integral representation:
\begin{equation}\footnotesize
	\frac{\Gamma(1+m+a\ep) \Gamma(1+n+b\ep)}{\Gamma(2+n+m+(a+b)\ep))}{ }_{2} \,F_{1}(-p-c\ep, 1+m+a\ep ; 2+n+m+(a+b)\ep) ; z)=\int_{0}^{1} d u\, u^{m+a\ep}(1-u)^{n+b\ep}(1-u \,z)^{p+c\ep}
\end{equation}
 We will consider $m$, $n$ and $p$ as integers. This renders the $\ep$ expansion expressible in MPLs. The homology and cohomology groups of the $\F$ are two-dimensional. The homology group is generated by considering contours whose endpoints are the zeroes of two of the polynomials appearing in the integral representation. In \cite{abreu2019diagrammatic} the contours that have been chosen are:
\begin{equation}\label{gamma}
	\bra{\gamma_{1}}=b \varepsilon\int_{0}^{1},\;\;\;\;\;\;	\bra{\gamma_{2}}=c \varepsilon z\int_{0}^{\frac{1}{z}}.
\end{equation}
The cohomology group is generated by two independent differential forms, given by:
\begin{equation}
	\ket{\omega_{1}}=d u\, u^{a \varepsilon}(1-u)^{-1+b \varepsilon}(1-u z)^{c \varepsilon},\;\;\;\;\;\;\;\;\;\;\;\;\;	\ket{\omega_{2}}=d u\, u^{a \varepsilon}(1-u)^{b \varepsilon}(1-u z)^{-1+c \varepsilon},
\end{equation}
such that they are dual to the above two contours. With these considerations the coaction of any $\F$ function is given by:
\begin{equation}\small
\Delta\left[\int_{0}^{1} d u\, u^{m+a\ep}(1-u)^{n+b\ep}(1-u \,z)^{p+c\ep}\right]=\Delta\left[\bra{\gamma}\ket{\omega}\right]=\bra{\gamma}\ket{\omega_1}\otimes\bra{\gamma_1}\ket{\omega}+\bra{\gamma}\ket{\omega_2}\otimes\bra{\gamma_2}\ket{\omega}.
\end{equation}

A variety of generalized hypergeometric functions and their coactions have been studied in \cite{Abreu:2019wzk}. In this work we will make use of the $\FF$ hypergeometric function, whose integral representation is given by:
\begin{equation}
	\ff{a_1}{a_2}{a_3}{b_1}{b_2}{z}=\frac{\G{b_2}}{\G{a_3}\G{b_2-a_3}}\int_{0}^{1}t^{a_3-1}\left(1-t\right)^{b_2-a_3-1}\f{a_1}{a_2}{b_1}{t\,z}.
\end{equation}
The construction of its coaction is very similar to the $\F$. Here the homology and cohomology spaces are three-dimensional.
\section{The on-shell double box top topology}
The subgroup of the cohomology group that includes the massless double box itself, without any of its lower subtopologies, is two-dimensional. As a result, there are two independent differential forms, and therefore two master integrands, that correspond to the double box top topology. 

In this section, we consider the differential form that spans the top topology, $\ket{\omega_{a,b,\{n_i=1\}}}=$\\$\ket{\omega_{a,b}}$, with $n_i=1$ for all $i$, and determine the values of $a$ and $b$ such that the duality condition of equation \eqref{dual} is realized, by computing the contours corresponding to the maximal cuts of the two double boxes. This process will determine the two master integrals of the double box.
\subsection{The maximal cuts of the double box and their coaction}
Our initial goal is to calculate the maximal cuts of the double box. The maximal cuts were first calculated in \cite{Bosma:2017ens} through the Baikov parametrization. We instead perform the calculation using an explicit phase-space parametrization of the cut loop momenta, as in \cite{SoutoGoncalvesdeAbreu:2015tfn}. Putting all the propagators on-shell, the resulting function is:
\begin{equation}\label{mc1}
\begin{split}
&\bra{\gamma_1}\ket{\omega_{a,b}}= f_{a,b}\;\f{1+2\ep}{b-\ep}{1+b-a}{x},
\end{split}
\end{equation}
with:
\begin{equation}\label{fab}
	f_{a,b}=-2e^{2\gamma_E\ep} \frac{\G{a-\ep}\G{-\ep}}{\G{-2\ep}\G{a-2\ep}}x^{-2-2\ep+b}(1-x)^{\ep}t^{a+b-3-2\ep}
\end{equation}
and where $x=-\frac{s}{t}$, $s=(p_1+p_2)^2$ and $t=(p_1+p_3)^2$.

 The contour of the $\F$ function of equation \eqref{mc1} corresponds to the $\bra{\gamma_{1}}$ generating contour of the $\F$ homology group \eqref{gamma}. Using the fact that the homology group of the $\F$ function is two-dimensional we may find a second maximal cut by restricting the integration range of the integral definition of the $\F$ to the $\bra{\gamma_2}$ contour of equation \eqref{gamma}. Doing so yields:
\begin{equation}\label{mc2}\small
		\bra{\gamma_2}\ket{\omega_{a,b}}=\frac{\Euler\G{a-\ep}\G{-\ep}\G{-\ep+b}}{\G{-2\ep}\G{a+b-3\ep}}t^{a+b-3-2\ep}x^{-2-\ep}(1-x)^{\ep}\f{b-\ep}{a-\ep}{a+b-3\ep}{\frac{1}{x}}
\end{equation}

We may then expand each result in $\ep$ to deduce the values of $a$ and $b$ that correspond to dual differential forms to the maximal cut contours, such that equation \eqref{dual} is satisfied. The result is: 
\begin{equation}
\begin{split}\label{a,b}
	&\bra{\gamma_1}\ket{\omega_{a=0,b=0}}=\bra{\gamma_1}\ket{\omega_{1}}=1+\mathcal{O}\left(\ep\right),\;\;\;\;\;\bra{\gamma_1}\ket{\omega_{a=0,b=1}}=\bra{\gamma_1}\ket{\omega_{2}}=\mathcal{O}(\ep)\\
	&\bra{\gamma_2}\ket{\omega_{a=0,b=0}}=\bra{\gamma_2}\ket{\omega_{1}}=\mathcal{O}\left(\ep\right),\,\,\,\,\,\;\,\,\,\,\,\,\,\;\;\;\;\bra{\gamma_2}\ket{\omega_{a=0,b=1}}=\bra{\gamma_2}\ket{\omega_{2}}=1+\mathcal{O}\left(\ep\right)
	\end{split}
\end{equation}
This choice coincides with the integrand basis of  \cite{Smirnov_1999} and \cite{Anastasiou_2000}, up to an overall normalization, and results in pure functions as shown in \cite{Henn:2013pwa}.

In terms of the its diagrammatic representation, we use different colors to represent the two master integrands: 
\begin{equation}\small
\begin{split}
\bra{\gamma}\ket{\omega_{a=0,b=0}}=\bra{\gamma}\ket{\omega_{1}}=\raisebox{-8.3mm}{\includegraphics[scale=0.305]{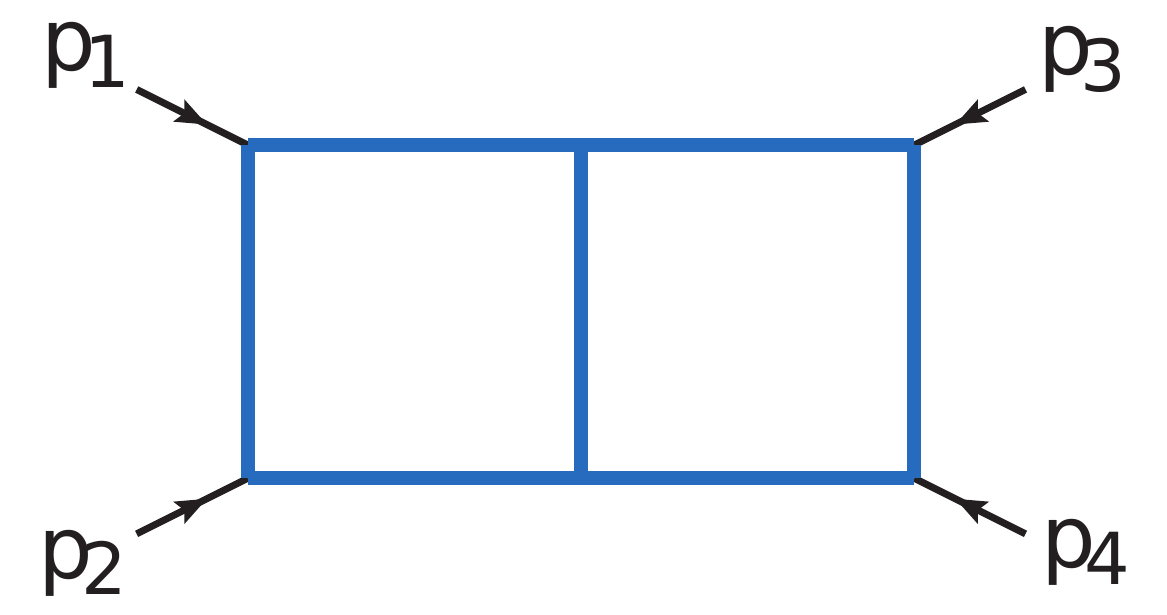}}\;\bra{\gamma}\ket{\omega_{a=0,b=1}}=\bra{\gamma}\ket{\omega_{2}}=\raisebox{-8.3mm}{\includegraphics[scale=0.305]{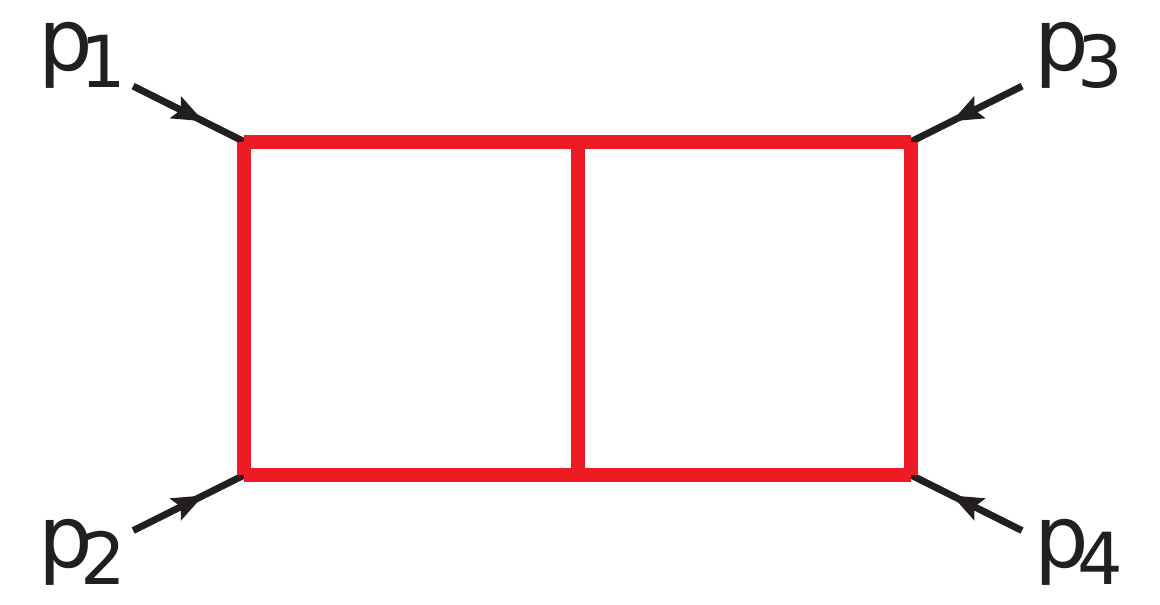}}
\end{split}
\end{equation}

Applying the cut contours to these master integrands, the space of the maximal-cut diagrams takes the form:
\begin{equation}\small
\begin{split}
\bra{\gamma_1}\ket{\omega_{1}}=\raisebox{-8.3mm}{\includegraphics[scale=0.305]{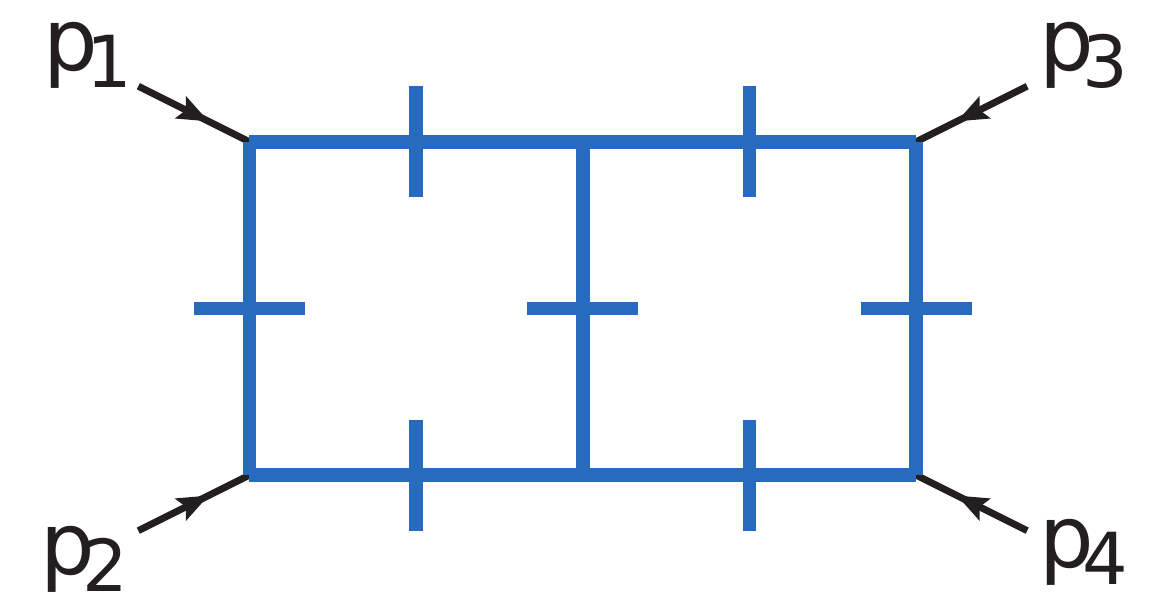}}&\;\;\;\;\;\;\;\;\;\;\;\;\;\;\;\;\;\;\;\;\;\;\;\;\;\;\;\;\bra{\gamma_1}\ket{\omega_{2}}=\raisebox{-8.3mm}{\includegraphics[scale=0.305]{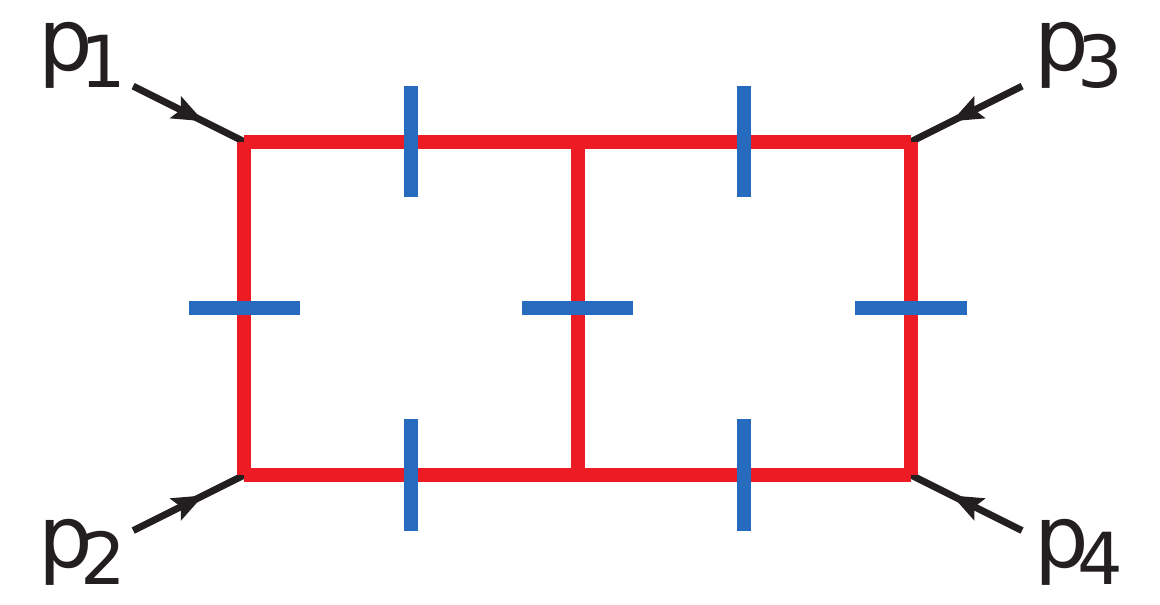}}\\
\bra{\gamma_2}\ket{\omega_{1}}=\raisebox{-8.3mm}{\includegraphics[scale=0.305]{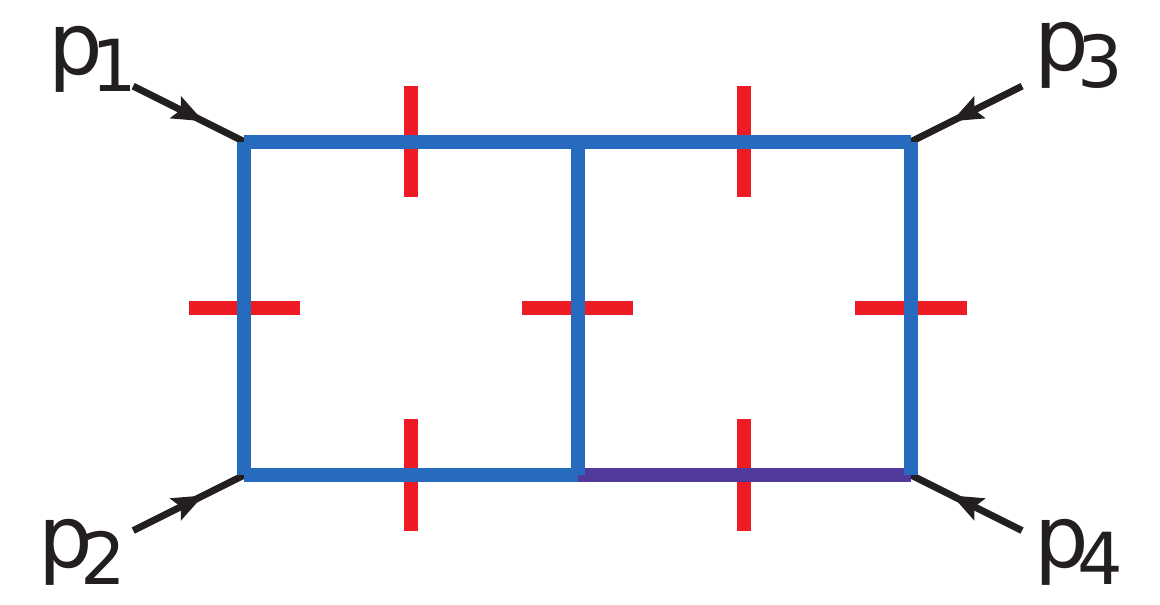}}&\;\;\;\;\;\;\;\;\;\;\;\;\;\;\;\;\;\;\;\;\;\;\;\;\;\;\;\;\bra{\gamma_2}\ket{\omega_{2}}=\raisebox{-8.3mm}{\includegraphics[scale=0.305]{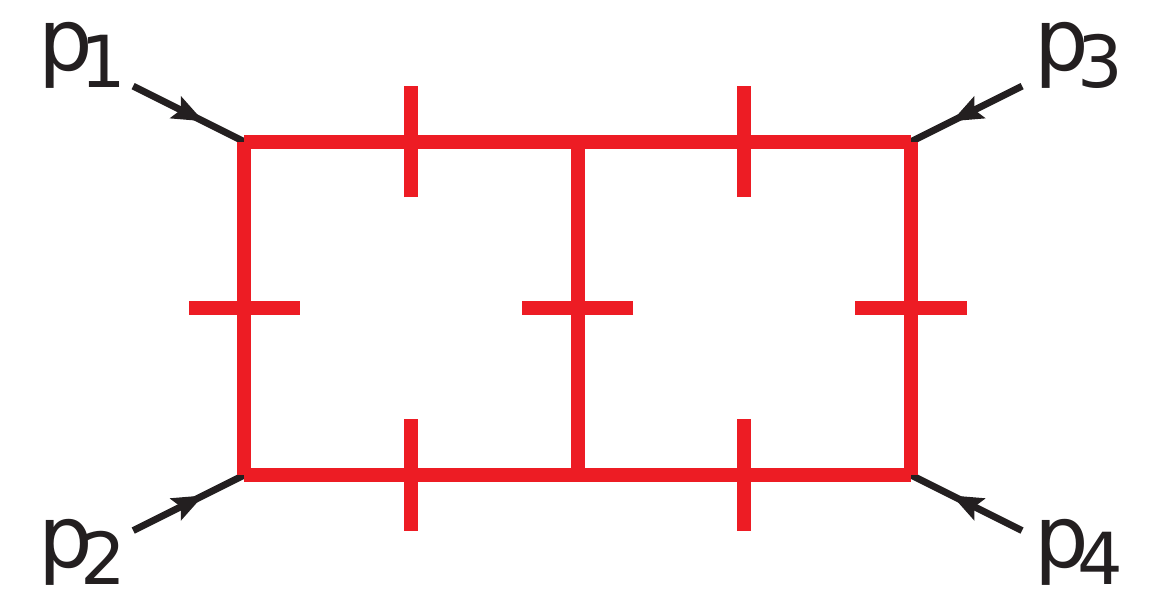}}
\end{split}
\end{equation}
Using the results of \cite{Abreu:2019wzk} we can calculate the coaction on the hypergeometric function corresponding to each of the maximal cuts and then identify the functions, in which the coaction is expressed, in terms of (cut) diagrams. In this way we obtain the form of the diagrammatic coaction on the maximal-cut subspace. For example, for $\bra{\gamma_1}\ket{\omega_{1}}$ we obtain:
	\begin{equation}
\Delta\left[\bra{\gamma_{1}}\ket{\omega_1}\right]=\bra{\gamma_{1}}\ket{\omega_1}\otimes\bra{\gamma_{1}}\ket{\omega_1}+\bra{\gamma_{1}}\ket{\omega_2}\otimes\bra{\gamma_{2}}\ket{\omega_1}
\end{equation}
 which is of the same form as equation \eqref{diagcoa} and can be represented diagrammatically as:
\begin{equation}\scriptsize
\begin{split}
\Delta\left[\raisebox{-5.8mm}{\includegraphics[scale=0.22]{DB1mc1.pdf}}\right]=\raisebox{-5.8mm}{\includegraphics[scale=0.22]{DB1mc1.pdf}}\otimes\raisebox{-5.8mm}{\includegraphics[scale=0.22]{DB1mc1.pdf}}+\raisebox{-5.8mm}{\includegraphics[scale=0.22]{DB2mc1.pdf}}\otimes \raisebox{-5.8mm}{\includegraphics[scale=0.22]{DB1mc2.pdf}}
\end{split}
\end{equation}
Here and below we write explicitly the results for the $\ket{\omega_{1}}$ (blue box) example. Similar results have been obtained for $\ket{\omega_{2}}$ (red box).
\section{The double box differential equations} \label{difsex}
\subsection{The homogenous equations}\label{homgeq}
We consider the effect of the differential operator $\frac{d}{dx}$ on the double-box maximal-cut subspace. Any subtopology with fewer propagators than those cut is then automatically set to zero. We may derive a differential equation using integration by parts (IBP) relations. IBP relations are the same on cut diagrams as on uncut ones.

 For the maximal cut corresponding to the $\gamma_1$ contour we obtain the following first-order differential equations:
 \begin{subequations}
\begin{equation}\label{difblue}
\frac{d}{dx}\raisebox{-6.9mm}{\includegraphics[scale=0.26]{DB1mc1.pdf}}=C_1(x,D)\raisebox{-6.9mm}{\includegraphics[scale=0.26]{DB1mc1.pdf}}+C_2(x,D)\raisebox{-6.9mm}{\includegraphics[scale=0.26]{DB2mc1.pdf}}
\end{equation}
\begin{equation}\label{difred}
\frac{d}{dx}\raisebox{-6.9mm}{\includegraphics[scale=0.26]{DB2mc1.pdf}}=	\tilde{C_1}(x,D)\raisebox{-6.9mm}{\includegraphics[scale=0.26]{DB1mc1.pdf}}+\tilde{C_2}(x,D)\raisebox{-6.9mm}{\includegraphics[scale=0.26]{DB2mc1.pdf}}
\end{equation}
\end{subequations}
where $C_i(x,D)$ and $\tilde{C_i}(x,D)$  $\left(\text{for}\, i=1,2\right)$ are rational functions of the spacetime dimension $D$ and the dimensionless ratio $x$. Focusing on the blue double box \eqref{difblue}, we may create a homogeneous differential equation by taking an extra derivative, thus increasing the order of the differential equation by one, and then eliminating the red double box, understood as an inhomogeneous term, using its first-order  differential equation \eqref{difred}. We obtain:
\begin{equation}\label{homog}
\begin{split}
\frac{d^2}{dx^2}&\raisebox{-6.9mm}{\includegraphics[scale=0.26]{DB1mc1.pdf}}+A(x,D)\frac{d}{dx}\raisebox{-6.9mm}{\includegraphics[scale=0.26]{DB1mc1.pdf}}+B(x,D)\raisebox{-6.9mm}{\includegraphics[scale=0.26]{DB1mc1.pdf}}=0
\end{split}
\end{equation}
where $A(x,D)$ and $B(x,D)$ are again rational functions of their arguments. The two independent solutions of the differential equation are given by the two  maximal cuts of the blue double box, given by equations \eqref{mc1} and \eqref{mc2} for $a,\,b=0$, by imposing the duality condition of equation \eqref{dual} as a boundary condition.
\subsection{Non-maximal cuts and inhomogeneous equations}\label{secinhomo}
In this section we consider a larger space of contours corresponding to non-maximal cuts. We compute next-to-next-to-maximal cuts of the double box as all next-to-maximal cuts satisfy the same differential equation as the maximal cuts, since all six propagator diagrams are reducible to diagrams with a lower number of propagators. Consequently, the next-to-maximal cuts are spanned by the maximal cuts and therefore are not part of our basis.

 Choosing a contour that puts fewer propagators on-shell increases the dimension of the  active subgroup of homology and cohomology. This can be seen in the form of the coaction of equation \eqref{diagcoa} where additional terms, associated with the non-maximal cuts, appear in the sum. In the differential equation, this is reflected by the existence of extra inhomogeneous terms.  Consider, for example, the contour, referred to as $\bra{\gamma_{3}}$, corresponding to the following five-propagator cut:
	\begin{equation}\label{g3rep}
\bra{\gamma_{3}}\ket{\omega_{1}}=\raisebox{-7.3mm}{\includegraphics[scale=0.28]{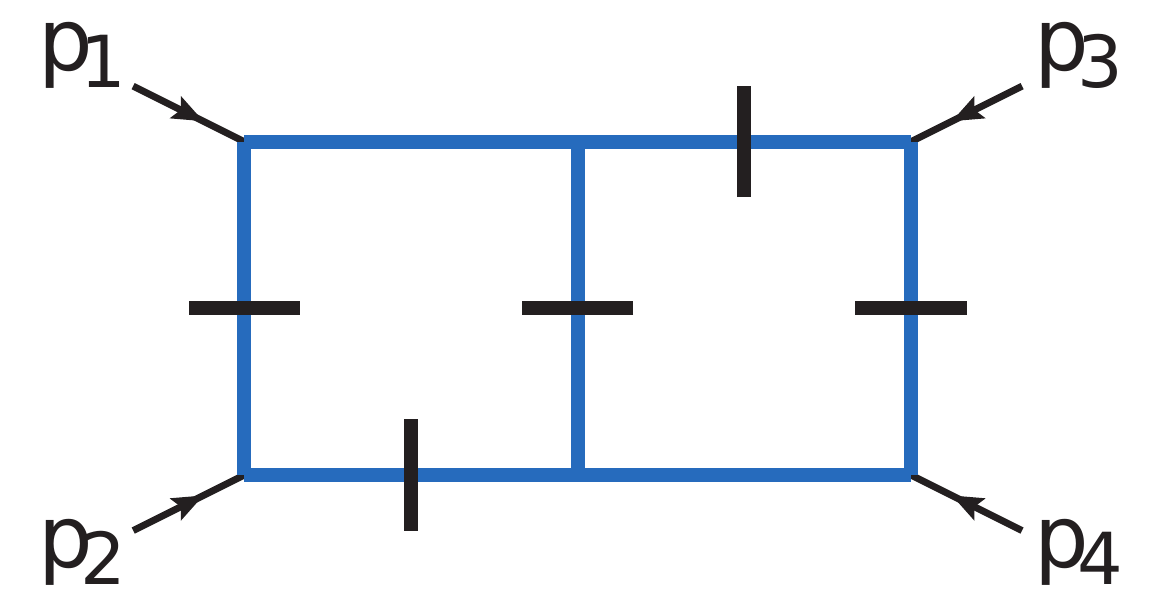}}.
\end{equation}
The differential equation on this diagram is:
\begin{equation}\label{inhomeq}\small
\begin{split}
\frac{d^2}{dx^2}&\raisebox{-5.8mm}{\includegraphics[scale=0.22]{DB1nnmc2.pdf}}+\, A(x,D)\frac{d}{dx}\raisebox{-5.8mm}{\includegraphics[scale=0.22]{DB1nnmc2.pdf}}+\, B(x,D)\raisebox{-5.8mm}{\includegraphics[scale=0.22]{DB1nnmc2.pdf}}=C(x,D)\raisebox{-5.8mm}{\includegraphics[scale=0.22]{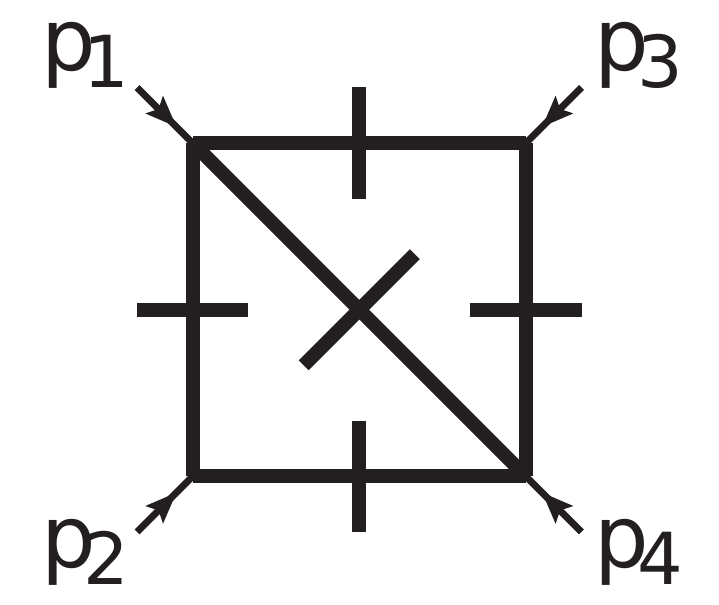}}
\end{split}
\end{equation}
where $A(x,D),\text{and}\,B(x,D)$ are the same coefficients as in equation \eqref{homog} and $C(x,D)$ is again a rational function. The diagonal box, appearing on the right hand side of the equation, has a single master integral associated with its topology and its maximal cut is known.

 This differential equation for the next-to-next maximal cut has the same homogeneous solution as equation \eqref{homog}. It also features a particular solution which corresponds to the inhomogenous term.
\paragraph{Solving the inhomogenous differential equations.}
To solve equation \eqref{inhomeq} we make use of the fact that maximal cut subspace is part of the solution space of this differential equation. Motivated from the form of equation \eqref{mc1} we define for $i=1,2,3$ :
\begin{equation}\label{normeq}
\begin{split}
&g_{i}(x)=\frac{\bra{\gamma_{i}}\ket{\omega_{1}}}{f_{a=0,b=0}},\qquad\text{with} \;g_1(x)=\f{-\ep}{1+2\ep}{1}{x}
\end{split}
\end{equation}
where $f_{a,b}$ has been defined in equation \eqref{fab}, $\bra{\gamma_i}$ for $i=1,2$ defined in equation \eqref{a,b} and $\bra{\gamma_3}$ corresponds to the cut contour represented in the diagram of equation  \eqref{g3rep} which we are interested in computing. The choice of normalization is motivated by the fact that it simplifies the form of $g_1(x)$, which is part of the space of solutions of the differential equation.

Starting with a first-order inhomogeneous differential equation for $g_3(x)$, we follow the procedure of section \ref{homgeq} to eliminate the diagonal-box inhomogeneous term and generate a third-order homogeneous differential equation for $g_{3}(x)$:
\begin{equation}\label{3rdhom}\small
	(1-x)x^2\frac{d^3g_{3}(x)}{dx^3}+2x(1-2x)\frac{d^2g_{3}(x)}{dx^2}+\left(x(3\ep^2+2\ep-2\ep)+4\ep(1+\ep)\right)\frac{dg_{3}(x)}{dx}-\ep^2(1+2\ep)g_{3}(x)=0
\end{equation}
This differential equation can be recognized as the defining differential equation of the $\FF$ function. The simplicity of its form follows directly from the normalization chosen in \eqref{normeq}. We can compare equation \eqref{3rdhom} with the general form of the $\FF$ differential equation \cite{Erdelyi:1953:HTF1} and identify the three-dimensional solution space of the cut diagram as:
\begin{equation}\label{solutions}
\begin{split}
\raisebox{-5.6mm}{\includegraphics[scale=0.22]{DB1nnmc2.pdf}}=&-2\frac{\G{-\ep}^2}{\G{-2\ep}^2}t^{-3-2\ep}x^{-2-2\ep}(1-x)^{\ep}\bigg[c_1(\ep)\,\f{-\ep}{1+2\ep}{1}{x}\\
+&c_2(\ep)\,\frac{(-x)^{-\ep}(1-x)^{-\ep}\G{1+3\ep}}{\G{1+\ep}\G{1+2\ep}}\f{1+2\ep}{1+2\ep}{2+3\ep}{\frac{1}{x}}\\
+&c_3(\ep)\,(1-x)^{1+2\ep} {}_3F_2\left(1,1,2+3 \ep;2+\ep,2+2\ep;1-x\right)\bigg]
\end{split}
	\end{equation}
	where $c_i(\ep),\;i=1,2,3$ are yet-undetermined coefficients which span the entire three-dimensional solution space. We note that the first two of the three terms span the maximal cut subspace of the double box and the third term is associated with the non-maximal cut.
	\paragraph{Determining the coefficients.}
Equation \eqref{solutions} can be understood as the sum of the homogeneous and particular solutions of equation \eqref{inhomeq}. As a result, the value of the $c_3(\ep)$ coefficient, which corresponds to the particular solution, is determined by the form of the diagonal-box inhomogeneous term. We can solve for $c_3(\ep)$ by generating a third-order differential equation which features the diagonal-box and demanding that \eqref{solutions} is a solution to the equation:
 \begin{equation}\label{coefeq}
\begin{split}
&\frac{d^3}{dx^3}\raisebox{-5.6mm}{\includegraphics[scale=0.22]{DB1nnmc2.pdf}}+C_1'\frac{d^2}{dx^2}\raisebox{-5.6mm}{\includegraphics[scale=0.22]{DB1nnmc2.pdf}}+\frac{d}{dx}C_2'\raisebox{-5.6mm}{\includegraphics[scale=0.22]{DB1nnmc2.pdf}}+C_3'\raisebox{-5.6mm}{\includegraphics[scale=0.22]{DiagBoxmc.pdf}}=0.
\end{split}
\end{equation}
Substituting \eqref{solutions} into \eqref{coefeq}, the terms that feature $c_1(\ep)$ and $c_2(\ep)$ vanish among themselves and we obtain an algebraic equation which determines $c_3(\ep)$:
 \begin{equation}
\begin{split}
&\scalebox{1.2}{$c_3=\frac{12\ep(1+3\ep)}{(1+\ep)(1+2\ep)}\frac{x^{2+2\ep}\left(1-x\right)^{1-2\ep}}{t^{-3-2\ep}}\raisebox{-6.5mm}{\includegraphics[scale=0.25]{DiagBoxmc.pdf}}=\frac{12\ep(1+3\ep)}{(1+\ep)(1+2\ep)}\frac{\G{1+3\ep}\G{1+2\ep}}{\G{1+\ep}^3}$},
\end{split}
\end{equation}
where in the last step we have inserted the maximal cut of the diagonal box, which we have separately computed using the Baikov parametrization.

To determine the values of $c_1(\ep)$ and $c_2(\ep)$, we again enforce the duality condition \eqref{dual} as a boundary condition. This requires to have solved the entire system of differential equations, which includes $\bra{\gamma_3}\ket{\omega_2}$ i.e., the same cut but applied to the red double box, obtained by setting up an equation similar to equation to \eqref{coefeq}, but for $\bra{\gamma_3}\ket{\omega_2}$. To implement the boundary condition we demand:
\begin{equation}
	\bra{\gamma_3}\ket{\omega_1}=\mathcal{O}(\ep),\qquad\bra{\gamma_3}\ket{\omega_2}=\mathcal{O}(\ep)
\end{equation}
which fixes $c_1(\ep)$ and $ c_2(\ep)$ to be:
\begin{equation}
	c_1(\ep)=0,\qquad c_2(\ep)=-\frac{3}{2}.
\end{equation}

The process of fixing the undetermined coefficients is equivalent to defining the $\bra{\gamma_3}$ contour. In this way we have determined the basis of the three dimensional cohomology group that includes the two maximal cuts as well as the $\bra{\gamma_3}$ cut.
\section{The coaction of cuts}
Having defined the $\bra{\gamma_3}$ contour we can now apply the coaction on the hypergeometric functions of $\bra{\gamma_3}\ket{\omega_1}$ and calculate the coaction of the cut, which takes the form of \eqref{diagcoa}:
\begin{equation}
	\Delta\left[\bra{\gamma_{3}}\ket{\omega_1}\right]=\bra{\gamma_{3}}\ket{\omega_1}\otimes\bra{\gamma_{1}}\ket{\omega_1}+\bra{\gamma_{3}}\ket{\omega_2}\otimes\bra{\gamma_{2}}\ket{\omega_1}+\bra{\gamma_{3}}\ket{\omega_3}\otimes\bra{\gamma_{3}}\ket{\omega_1}
\end{equation}
Diagrammatically this takes the form:
\begin{equation}\scriptsize
\begin{split}
		\Delta\left[\raisebox{-5.8mm}{\includegraphics[scale=0.22]{DB1nnmc2.pdf}}\right]&=\raisebox{-5.8mm}{\includegraphics[scale=0.22]{DB1nnmc2.pdf}}\otimes\raisebox{-5.8mm}{\includegraphics[scale=0.22]{DB1mc1.pdf}}+\raisebox{-5.8mm}{\includegraphics[scale=0.22]{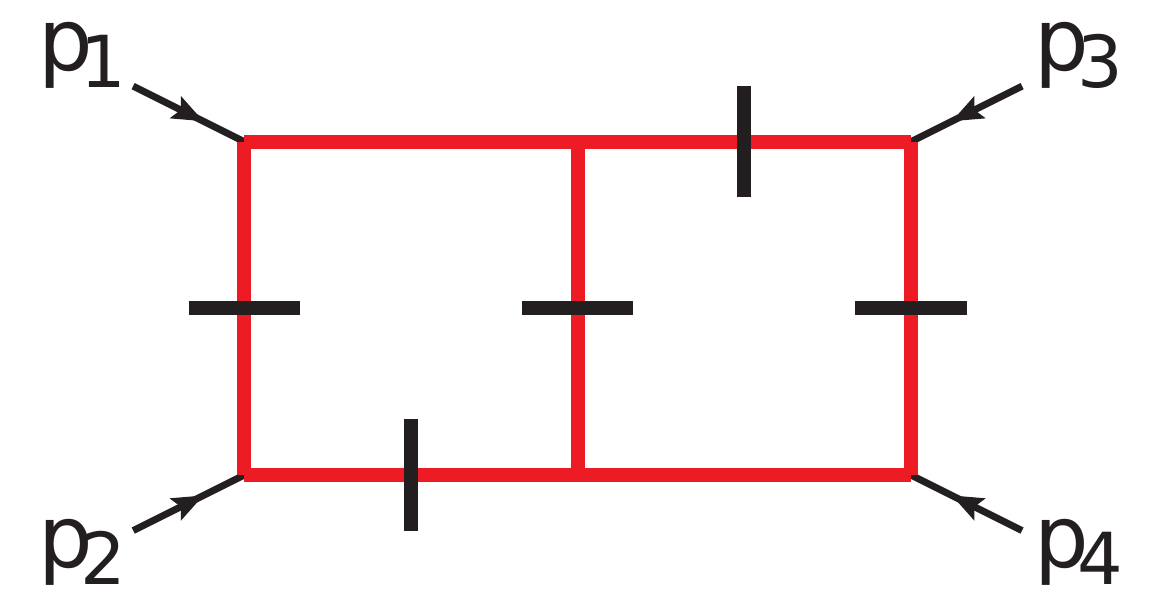}}\otimes\raisebox{-5.8mm}{\includegraphics[scale=0.22]{DB1mc2.pdf}}\\
		&+\raisebox{-5.8mm}{\includegraphics[scale=0.22]{DiagBoxmc.pdf}}\otimes\raisebox{-5.8mm}{\includegraphics[scale=0.22]{DB1nnmc2.pdf}}.
	\end{split}
\end{equation}

We have also followed the procedure outlined in section \ref{secinhomo} to define the $\bra{\gamma_4}$ and $\bra{\gamma_5}$ contours, corresponding to the following cut diagrams:
\begin{equation}
\bra{\gamma_{4}}\ket{\omega_1}=\raisebox{-5.8mm}{\includegraphics[scale=0.23]{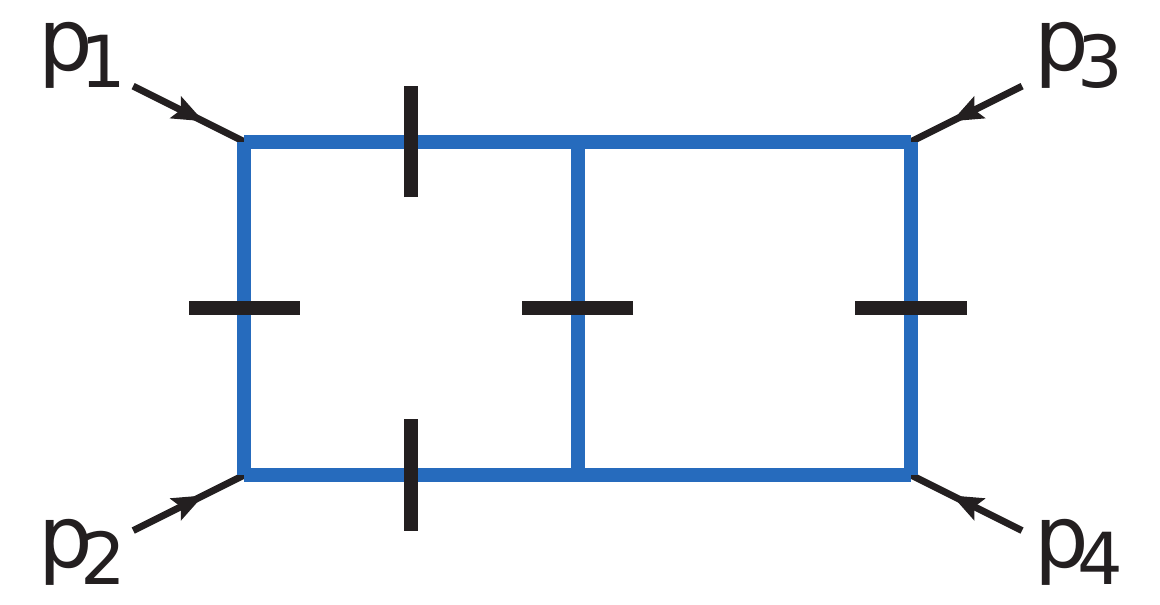}},\qquad\;\bra{\gamma_{5}}\ket{\omega_1}=\raisebox{-5.8mm}{\includegraphics[scale=0.23]{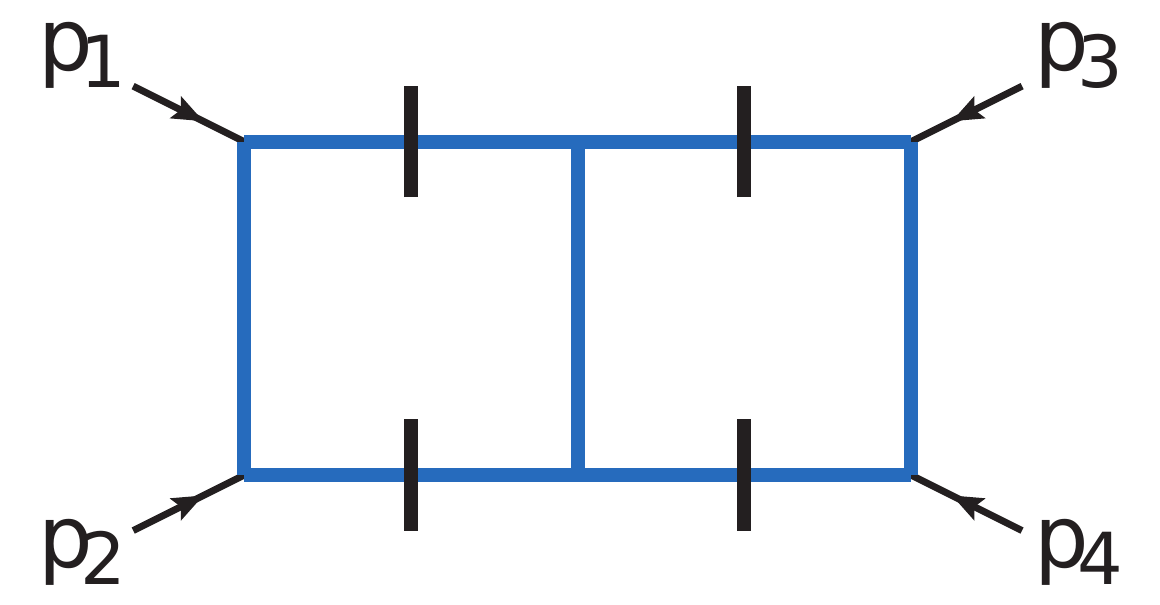}}.
\end{equation}
The spaces of these cuts are also three dimensional, containing the two maximal cuts as well the non-maximal cut. The diagrammatic coaction on these cuts of the double double box is given by:
\begin{equation}\scriptsize
\begin{split}
\Delta\left[\raisebox{-5.4mm}{\includegraphics[scale=0.22]{DB1nnmc1.pdf}}\right]&=\raisebox{-5.4mm}{\includegraphics[scale=0.22]{DB1nnmc1.pdf}}\otimes\raisebox{-5.4mm}{\includegraphics[scale=0.22]{DB1mc1.pdf}}+\raisebox{-5.4mm}{\includegraphics[scale=0.22]{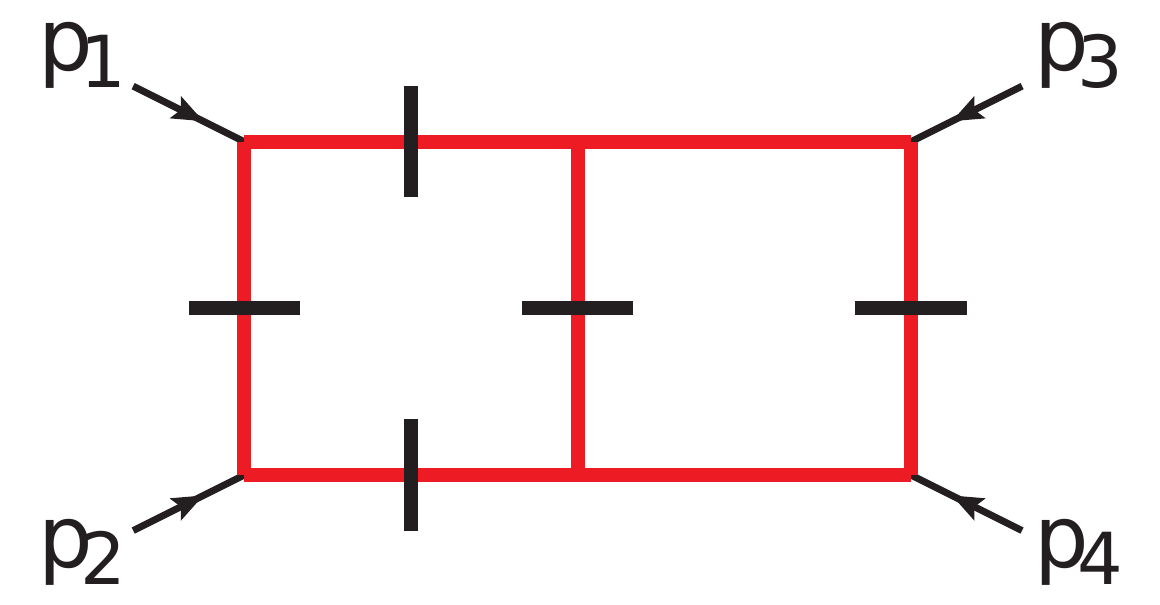}}\otimes\raisebox{-5.4mm}{\includegraphics[scale=0.22]{DB1mc2.pdf}}\\
&+\raisebox{-5.4mm}{\includegraphics[scale=0.22]{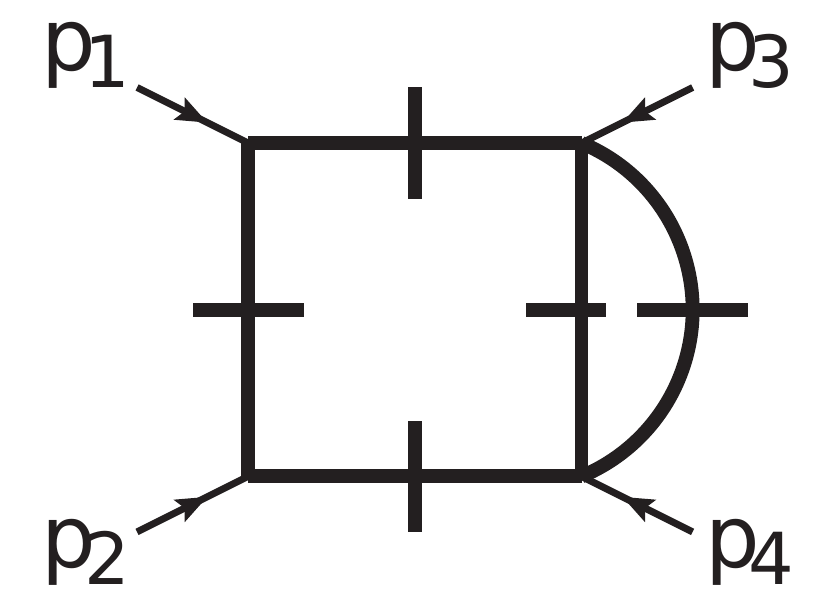}}\otimes\raisebox{-5.4mm}{\includegraphics[scale=0.22]{DB1nnmc1.pdf}}
\end{split}
\end{equation}
\begin{equation}\scriptsize
\begin{split}
	\Delta\left[\raisebox{-5.4mm}{\includegraphics[scale=0.22]{DB1nnnmc1.pdf}}\right]&=\raisebox{-5.4mm}{\includegraphics[scale=0.21]{DB1nnnmc1.pdf}}\otimes\raisebox{-5.4mm}{\includegraphics[scale=0.22]{DB1mc1.pdf}}+\raisebox{-5.4mm}{\includegraphics[scale=0.22]{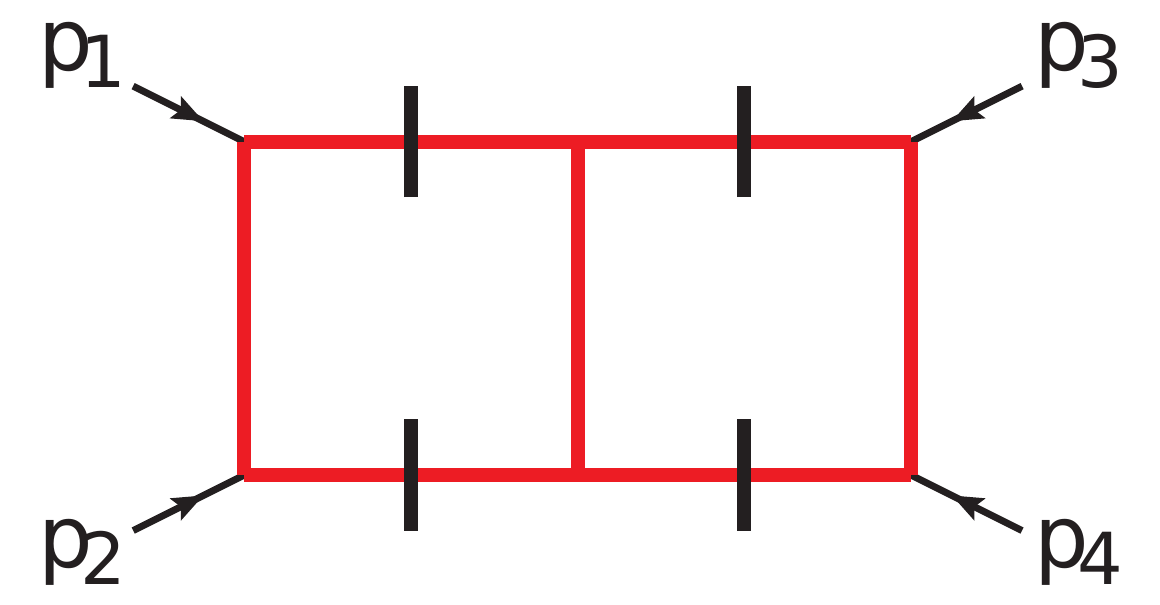}}\otimes\raisebox{-5.4mm}{\includegraphics[scale=0.22]{DB1mc2.pdf}}\\
&+\raisebox{-3.8mm}{\includegraphics[scale=0.25]{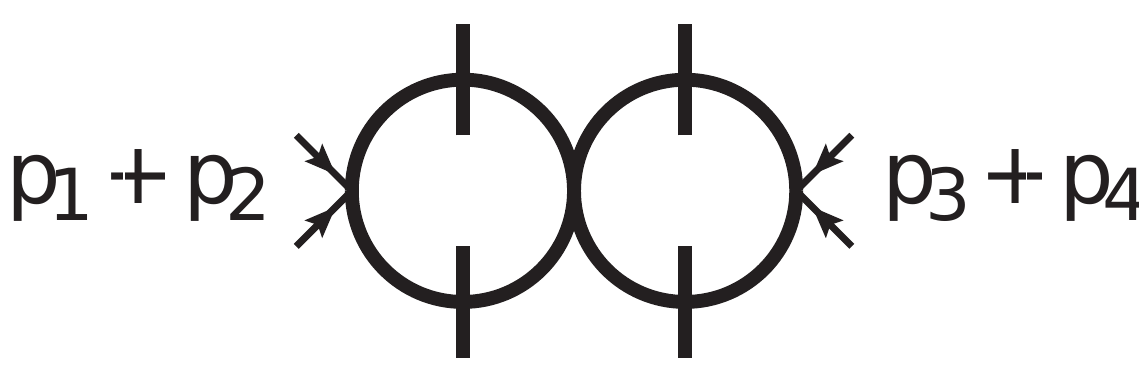}}\,\otimes\raisebox{-5.4mm}{\includegraphics[scale=0.22]{DB1nnnmc1.pdf}}.
\end{split}
\end{equation}
\section{Summary and discussion}
 We have calculated the coaction for a number of cuts of the double box, defining the cohomology group basis associated with each cut. We have also demonstrated how the solution of the differential equation for each cut appears in the coaction and how the duality condition, required by the coaction, can be used as a boundary condition.
 
  The differential equations for this work were obtained using IBP relations. A non trivial consistency check to be performed is to derive the same differential equations from the coaction. Constructing differential equations in this way allows one to interpret the coefficients of the differential equations via the $\ep$ expansion of the cut diagrams appearing in the coaction \cite{Abreu:2017mtm}. Future research will also focus on calculating the remaining cuts of the double box and thus further testing the differential-equation techniques discussed here.
  
 While the form of the two-loop coaction in the form of \eqref{diagcoa} has been well established, an algorithmic approach to choosing the basis of integration contours that satisfy \eqref{diagcoa}, is yet to be developed. This has been achieved at the one-loop level in full generality. Analyzing the coaction on the uncut double box topology will contribute towards this goal at the two-loop level.
 
 A related aspect of these calculations, not discussed in the present talk, is the definition of explicit integration contours by identifying specific integration endpoints for every parameter of a given parametrization, similarly to the work conducted in \cite{Bosma:2017ens} and \cite{Harley:2017qut} for the Baikov parametrization. In the context of non-maximal double box cuts, these direct integration techniques are harder to use for obtaining results in a closed form in $\ep$, but are nevertheless an important aspect to consider in order to determine a general rule for the basis of integration contours of two-loop diagrams, required by the coaction.

\section*{Acknowledgements}
We would like to thank S. Abreu, R. Britto and C. Duhr for many interesting discussions and helpful comments.

% TODO: include funding information
\paragraph{Funding information}
This work has been funded in part by the STFC Consolidated Grant “Particle Physics at the Higgs Centre".

\bibliography{The_diagrammatic_coaction_and_cuts_of_the_double_box.bib}
\nolinenumbers

\end{document}